\documentclass[11pt]{article}
\usepackage{graphicx}
\begin{document}
\title{Use of Nilpotent weights in Logarithmic Conformal Field Theories \footnote{
Talk delivered in school and workshop on Logarithmic Conformal
Field Theory, Tehran, Iran, 2001}}
\author{S. Moghimi-Araghi \footnote{e-mail: samanimi@rose.ipm.ac.ir} ,
S. Rouhani \footnote{e-mail: rouhani@ipm.ir} and M.
Saadat \footnote{e-mail: saadat@mehr.sharif.ac.ir}\\
\\
Department of Physics, Sharif University of Technology,\\ Tehran,
P.O.Box: 11365-9161, Iran\\ Institute for studies in Theoretical
physics and Mathematics,\\ Tehran, P.O.Box: 19395-5531, Iran}
\date{}
\maketitle
\begin{abstract}
We show that logarithmic conformal field theories may be derived
using nilpotent scale transformation. Using such nilpotent weights
we derive properties of LCFT's, such as two and three point
correlation functions solely from symmetry arguments. Singular
vectors and the Kac determinant may also be obtained using these
nilpotent variables, hence the structure of the four point
functions can also derived. This leads to non homogeneous
hypergeometric functions. We can construct "superfields" using a
nilpotent variable. Using this construct we show that the
superfield of conformal weight zero, composed of the identity and
the pseudo identity is related to a superfield of conformal
dimension two, which comprises of energy momentum tensor and its
logarithmic partner. This device also allows us to derive the
operator product expansion for logarithmic operators. Finally we
consider LCFT's near a boundary.
\\

\end{abstract}
\section{Introduction}
Logarithmic conformal field theories (LCFT) were first introduced
by Gurarie \cite{gur} in the context of $c=-2$ conformal field
theory (CFT). The difference between an LCFT and a CFT
\cite{bpz}, lies in the appearance of logarithmic as well powers
in the singular behavior of the correlation functions. In an LCFT,
degenerate groups of operators may exist which all have the same
conformal weight. They form a Jordan cell under the action of
$L_{0}$. In the simplest case a pair of operators exist which
transform according to
\begin{eqnarray}\label{eq:a1}
\phi(\lambda z)&=&\lambda^{-\Delta}\phi(z) ,\nonumber\\
\psi(\lambda z)&=&\lambda^{-\Delta}[\psi(z)-\phi(z)\ln \lambda] .
\end{eqnarray}
The correlation function of this pair were derived in
\cite{cau,rah,ghez}. Now using nilpotent variables \cite{mogh}
\begin{eqnarray}\label{eq:a2}
\theta_{i}^{2}&=&0 ,\nonumber\\
\theta_{i}\theta_{j}&=&\theta_{j}\theta_{i},
\end{eqnarray}
and  postulating that the conformal weights may have a  nilpotent
part
\begin{eqnarray}\label{eq:a3}
\Phi(\lambda z,\theta)&=&\lambda^{-(\Delta+\theta)}\Phi(z,\theta),
\end{eqnarray}
and the construct $\Phi(z,\theta)=\phi(z)+\theta \psi(z)$, we
arrive at equation (\ref{eq:a1}). In order to construct bigger
Jordan cells it is sufficient to expand our construct, using
$\theta^{n}=0$ and
\begin{eqnarray}\label{eq:a4}
\Phi(z,\theta)&=&
\phi_{0}(z)+\phi_{1}(z)\theta+\phi_{2}(z)\theta^{2}+\ldots+\phi_{n-1}(z)\theta^{n-1}.
\end{eqnarray}
We will however restrict ourselves to rank two cell, and
generalization to higher order cells is straight forward.
\section{Two and three point correlation functions}
Now consider the following two point function
\begin{eqnarray}\label{eq:a5}
G(z_{1},z_{2},\theta_{1},\theta_{2})=
\left<\Phi_{1}(z_{1},\theta_{1})\Phi_{2}(z_{2},\theta_{2})\right>,
\end{eqnarray}
invariance under translation and rotation force dependence of
$G(z_{1},z_{2},\theta_{1},\theta_{2})$ to be on $z_{1}-z_{2}$.
Under scale transformation we have
\begin{eqnarray}\label{eq:a6}
G(\lambda(z_{1}-z_{2}),\theta_{1},\theta_{2})&=&
\lambda^{-(\Delta_{1}+\theta_{1})}\lambda^{-(\Delta_{2}+
\theta_{2})}G((z_{1}-z_{2}),\theta_{1},\theta_{2}) ,
\end{eqnarray}
and under special conformal transformation
$z\longrightarrow\frac{z}{1+bz}$ we find that
$\Delta_{1}=\Delta_{2}=:\Delta$ and constant term in the
expansion of $G(z_{1},z_{2},\theta_{1},\theta_{2})$ vanishes. Then
\begin{eqnarray}\label{eq:a7}
\left<\Phi(z_{1},\theta_{1})\Phi(z_{2},\theta_{2})\right>&=&
\frac{1} {(z_{1}-z_{2})^{2\Delta+(\theta_{1}+\theta_{2})}}
(a_{1}(\theta_{1}+\theta_{2})+a_{12}\theta_{1}\theta_{2}).
\end{eqnarray}
Expanding both sides in terms of $\theta_{1}$ and $\theta_{2}$
leads to all possible correlation functions including $\phi$ and
$\psi$
\begin{eqnarray}\label{eq:a8}
\left<\phi(z^{\prime})\phi(z)\right>&=&0 ,\nonumber\\
\left<\psi(z^{\prime})\psi(z)\right>&=
&\frac{1}{(z^{\prime}-z)^{2\Delta}}(-2a_{1}\ln(z^{\prime}-z)+a_{12}) ,\nonumber\\
\left<\psi(z^{\prime})\phi(z)\right>&=&\frac{a_{1}}{(z^{\prime}-z)^{2\Delta}}
.
\end{eqnarray}
which is consistent with previous works \cite{cau}.

Let us next consider the three point function
\begin{eqnarray}\label{eq:a9}
G(z_{1},z_{2},z_{3},\theta_{1},\theta_{2},\theta_{3})&=&\left<
\Phi_{1}(z_{1},\theta_{1})\Phi_{2}(z_{2},\theta_{2})
\Phi_{3}(z_{3},\theta_{3})\right>.
\end{eqnarray}
Again it is clear that if one obtains this correlation function,
all the correlators such as
$\langle\phi_{1}\phi_{2}\phi_{3}\rangle$,
$\langle\phi_{1}\phi_{2}\psi_{3}\rangle$, $\cdots$ can be
calculated readily by expanding this correlation function in
terms of $\theta_{1}$, $\theta_{2}$ and $\theta_{3}$. Also note
that these fields may belong to different Jordan cells. The
procedure of finding this correlator is just the same as the one
we did for the two point function. Like ordinary CFTs, the three
point function is obtained up to some constants. Of course, in
our case it is found up to a function of $\theta_{i}$'s, {\it
i.e.}
\begin{eqnarray}\label{eq:a10}
G(z_{1},z_{2},z_{3},\theta_{1},\theta_{2},\theta_{3})&=&
f(\theta_{1},\theta_{2},\theta_{3})z_{12}^{-a_{12}}z_{23}^{-a_{23}}z_{31}^{-a_{31}},
\end{eqnarray}
where $z_{ij}=(z_{i}-z_{j})$ and
$a_{ij}=\Delta_{i}+\Delta_{j}-\Delta_{k}+(\theta_{i}+\theta_{j}-\theta_{k})$.
There are some constraints on
$f(\theta_{1},\theta_{2},\theta_{3})$, but further reduction
requires specification the rank of Jordan cell. As we have taken
it to be 2, we have
\begin{eqnarray}\label{eq:a11}
f(\theta_{1},\theta_{2},\theta_{3})=\sum_{i=1}^{3}C_{i}\theta_{i}
+\sum_{1\leq i<j\leq3}C_{ij}\theta_{i}\theta_{j}
+C_{123}\theta_{1}\theta_{2}\theta_{3} .
\end{eqnarray}
While symmetry considerations do not rule out a constant term on
the right hand side of equation (\ref{eq:a11}), but a consistent
OPE forces this constant to vanish \cite{mog,floh2}. This form
together with the equations (\ref{eq:a10}) and (\ref{eq:a11})
leads to correlation functions already obtained in the
literature. This is also consistent with the observation that in
all the known LCFT's so far the three point function of the first
field in the Jordan cell vanishes.

In a similar fashion one can derive the form of the four point
functions. But before this is done, we need to address the
question of singular vectors in an LCFT.
\section{Hilbert Space}
Considering the infinitesimal transformation consistent with
equation (\ref{eq:a3}) we have
\begin{eqnarray}\label{eq:a12}
\delta\Phi&=&-\epsilon\partial\Phi-(\Delta+\theta)\Phi\partial\epsilon.
\end{eqnarray}
This defines the action of the generators of the Virasoro algebra
on the primary fields and points to the existence of a highest
weight vector with nilpotent eigenvalue
\begin{eqnarray}\label{eq:a13}
L_{0}|\Delta+\theta\rangle&=&(\Delta+\theta)|\Delta+\theta\rangle ,\nonumber\\
L_{n}|\Delta+\theta\rangle&=&0,\:\:\:\:\:\:\:\:\:n\geq1\:\:.
\end{eqnarray}
Nilpotent state $|\Delta+\theta\rangle$ can be considered as
\begin{eqnarray}\label{eq:a14}
|\Delta+\theta\rangle&=&\Phi(0,\theta)|0\rangle=[\phi(0)+\theta\psi(0)]|0\rangle,\nonumber\\
&=&|\Delta,0\rangle+ \theta |\Delta,1 \rangle .
\end{eqnarray}
It can be easily seen that the law written in equation
(\ref{eq:a14}), leads to the well known equations
\begin{eqnarray}\label{eq:a15}
L_{0}|\Delta,0\rangle=\Delta|\Delta,0\rangle ,\nonumber\\
L_{0}|\Delta,1\rangle=\Delta|\Delta,1\rangle+|\Delta,0\rangle.
\end{eqnarray}
Now we define an out state in an LCFT
\begin{eqnarray}\label{eq:a16}
\langle \Delta+\theta|&=&\langle 0|\Phi^{\dag}(0,\theta),\nonumber\\
&=&\langle 0|[\phi(0)+\theta\psi(0)] ^{\dag},\nonumber\\
\end{eqnarray}
where dagger means adjoint of fields, just the same as CFT. So
\begin{equation}\label{eq:a17}
\langle \Delta+\theta|=\lim_{\acute{z}\rightarrow \infty}\langle
0|\left({\phi(\acute{z})
\acute{z}^{2\Delta}+\bar{\theta}[\psi(\acute{z})+\ln\acute{z}^2\phi(\acute{z})]}\acute{z}^{2\Delta}\right),
\end{equation}
which together 'in' state defined in equation (\ref{eq:a14}) and
using form of the two point correlation functions in equation
(\ref{eq:a8}) leads to
\begin{eqnarray}\label{eq:a18}
\langle \Delta+\theta|\Delta+\theta\rangle=\theta+
\bar{\theta}+d\:\: \bar{\theta}\theta,
\end{eqnarray}
where $d$ is $a_{12}$ if we normalize $a_{1}$ to one. By expanding
left hand side of the last equation, we find
\begin{equation}\label{eq:a19}
\langle \Delta,0|\Delta,0\rangle=0,\:\:\:\:\langle
\Delta,0|\Delta,1\rangle=\langle
\Delta,1|\Delta,0\rangle=1,\:\:\:\:\langle
\Delta,1|\Delta,1\rangle=d.
\end{equation}
In addition to these highest weight states, there are descendants
which can be obtained by applying $L_{-n}$'s on the highest
weight vectors
\begin{eqnarray}\label{eq:a20}
|\Delta+n_{1}+n_{2}+\cdots +n_{k}+\theta\rangle &=&
L_{-n_{1}}L_{-n{2}}\cdots L_{-n_{k}}|\Delta+\theta\rangle.
\end{eqnarray}
Most general state of Hilbert space at level $n$ can be
constructed as follows
\begin{eqnarray}\label{eq:a21}
|\chi_{\Delta,c}^{(n)}(\theta)\rangle&=&\sum_{|\vec{n}|=n}b^{\vec{n}}L_{-\vec{n}}|\Delta+\theta\rangle,\nonumber\\
&=&\sum_{\{n_{1}+n_{2}+\ldots+n_{k}=n\}} b^{(n_{1},n_{2},\cdots,
n_{k})}L_{-n_{1}}L_{-n_{2}}\ldots,
L_{-n_{k}}|\Delta+\theta\rangle.
\end{eqnarray}
As an application of the above definitions we compute  the
character formula
\begin{eqnarray}\label{eq:a22}
\chi_{\Delta}(\theta,\bar{\theta})&=&\sum_{N}\langle
N+\Delta+\theta| \eta^{L_{0}-\frac{c}{24}}|N+\Delta+\theta\rangle,
\end{eqnarray}
which by equation (\ref{eq:a20}) simplifies to
\begin{eqnarray}\label{eq:a23}
\chi_{\Delta}(\theta,\bar{\theta})&=&\eta^{\Delta+\theta-\frac{c}{24}}
\sum_{N} \eta^{N}p(N,\theta)\langle \Delta+\theta
|\Delta+\theta\rangle .
\end{eqnarray}
Writing $p(N,\theta)=p_{0}(N)+\theta p_{1}(N)$ we obtain four
characters
\begin{eqnarray}\label{eq:a24}
\chi_{\Delta}^{(\phi,\phi)}&=&0 ,\nonumber\\
\chi_{\Delta}^{(\phi,\psi)}&=&\chi_{\Delta}^{(\psi,\phi)}=
\eta^{\Delta-\frac{c}{24}}\sum_{N} \eta^{N}p_{0}(N) ,\nonumber\\
 \chi_{\Delta}^{(\psi,\psi)}&=&\eta^{\Delta-\frac{c}{24}}\sum_{N}
\eta^{N}\left[p_{1}(N)+(d+\ln\eta)p_{0}(N)\right] .
\end{eqnarray}
Appearance of logarithms in character formula have been discussed
in \cite{floh1,kog}.
\section{Singular Vectors in LCFT}
We define a singular vector at level $n$ by
$|\chi^{(n)}_{\Delta,c}\rangle_{S}$ as a vector that is
orthogonal to all vectors in its level
\begin{equation}\label{eq:a25}
\langle\chi^{(n)}_{\Delta^{\prime},c}|\chi^{(n)}_{\Delta,c}\rangle_{S}=0.
\end{equation}
As a result it has a zero norm and is orthogonal to any vector at
higher levels. According to equation (\ref{eq:a20}), last
condition is equivalent to
\begin{equation}\label{eq:a26}
\langle
\Delta^{'}+\theta|L_{\vec{n}^{'}}|\chi^{(n)}_{\Delta,c}\rangle_{S}=0,\hspace{1cm}\forall\:\:\:\vec{n}^{'}:
|\vec{n}^{'}|=n.
\end{equation}
The number of $L_{\vec{n}^{'}}$'s is $p(n)$, the number of
partitions of the integer $n$. So equation (\ref{eq:a26}) is
equivalent to $p(n)$ equation with $p(n)$ unknown coefficients. By
putting equation (\ref{eq:a21}) in the last equation we have
\begin{equation}\label{eq:a27}
\langle
\Delta^{'}+\theta|L_{\vec{n}^{'}}|\chi^{(n)}_{\Delta,c}\rangle_{S}=\sum_{|\vec{n}|=n}
b^{\vec{n}}\langle
\Delta^{'}+\theta|L_{\vec{n}^{'}}L_{-\vec{n}}|\Delta+\theta\rangle=0,
\hspace{1cm}\forall\:\:\:\vec{n}^{'}: |\vec{n}^{'}|=n,
\end{equation}
Since $L_{k}|\Delta+\theta\rangle=0$ (for $k\geq1$) and using
Virasoro algebra we find
\begin{equation}\label{eq:a28}
L_{\vec{n}^{'}}L_{-\vec{n}}|\Delta+\theta\rangle=
\sum_{m=0}\alpha_{m}^{\vec{n}^{'},\vec{n}}(c)L_{0}^{m}|\Delta+\theta\rangle,
\end{equation}
where coefficients $\alpha_{m}^{\vec{n}^{'},\vec{n}}(c)$ are
numbers or constants dependent on central charge $c$. By putting
this expression in equation (\ref{eq:a27})
\begin{equation}\label{eq:a29}
\sum_{|\vec{n}|=n}b^{\vec{n}}\left(
\sum_{m=0}\alpha_{m}^{\vec{n}^{'},\vec{n}}(c)(\Delta+\theta)^{m}\right)
\langle
\Delta^{'}+\theta|\Delta+\theta\rangle=0,\hspace{1cm}\forall\:\:\:\vec{n}^{'}:
|\vec{n}^{'}|=n,
\end{equation}
where $(\Delta+\theta)^{m}$ is eigenvalue of
$L_{0}^{m}|\Delta+\theta\rangle$. Non zero solutions for
$b^{\vec{n}}$'s leads to Kac determinant in LCFT
\begin{equation}\label{eq:a30}
\det\left(
\sum_{m=0}\alpha_{m}^{\vec{n}^{'},\vec{n}}(c)(\Delta+\theta)^{m}\right)=0\hspace{1cm}\forall\:\:\:|\vec{n}^{'}|=n,\:\:
|\vec{n}|=n.
\end{equation}
Singular vectors exist for those values of $\Delta$ and $c$ that
Kac determinant is zero just the same as CFT.

In the following we determine the null vectors at level 2 for a
Jordan cell of rank 2. We thus have
\begin{eqnarray}\label{eq:a31}
|\chi_{\Delta,c}^{(2)}(\theta)\rangle_{S}&=&\left(b^{(1,1)}L_{-1}^{2}+b^{(2)}L_{-2}\right)|\Delta+\theta\rangle.
\end{eqnarray}
Equation (\ref{eq:a29}) is equivalent to
\begin{equation}\label{eq:a32}
\left(\begin{array}{ccccc}
       &4(2\Delta^{2}+\Delta)+4(4\Delta+1)\theta&&6(\Delta+\theta)&\\
       &6(\Delta+\theta)&&4(\Delta+\theta)+\frac{c}{2}&
\end{array}\right)\left(\begin{array}{c}
b^{(1,1)}\\b^{(2)}\end{array}\right)=0,
\end{equation}
which leads to
\begin{equation}\label{eq:a33}
\Delta\left[8\Delta^{2}-5\Delta+c(\Delta+\frac{1}{2})\right]+\left[24\Delta^{2}-10\Delta+c(2\Delta+\frac{1}{2})\right]\theta=0.
\end{equation}
So Kac determinant vanishes for
$(\Delta,c)=(0,0),(-\frac{5}{4},25),(\frac{1}{4},1)$. Because of
homogeneity of the equation (\ref{eq:a32}) at least one of the
coefficients is arbitrary. Therefore we write
\begin{equation}\label{eq:a34}
b^{(1,1)}=3, \hspace{2cm} b^{(2)}=-[4(\Delta+\theta)+2],
\end{equation}
and since for a Jordan cell of rank 2,
$|\chi_{\Delta,c}^{(2)}(\theta)\rangle_{S}=|\chi_{\Delta,c}^{(2)}(0)\rangle_{S}+\theta|\chi_{\Delta,c}^{(2)}(1)\rangle_{S}$
we find
\begin{eqnarray}\label{eq:a35}
|\chi_{\Delta,c}^{(2)}(1)\rangle_{S}=
\left[3L_{-1}^{2}-2(2\Delta+1)L_{-2}\right]|\Delta,1\rangle-4L_{-2}|\Delta,0\rangle,
\end{eqnarray}
for $\Delta=\frac{1}{4},-\frac{5}{4}$ as a logarithmic singular
vector. $\Delta=0$ does not lead to a logarithmic singular vector
because $|\Delta,1\rangle$ does not appear in
$|\chi_{\Delta,c}^{(2)}(1)\rangle_{S}$. By the same technique
logarithmic singular vectors can be obtained at higher levels
which is consistent with \cite{floh3}.
\section{Roots of Kac determinant in LCFT}
In the previous section we saw that in an LCFT singular vectors
exist for those values of $(\Delta,c)$ which Kac determinant
vanishes just the same as CFT. If we compare equations
(\ref{eq:a28}) and (\ref{eq:a30}) with their counterparts in CFT,
we see that $\Delta$ in CFT has been replaced by $\Delta+\theta$.
One thus concludes that in an LCFT and at level $n$ Kac
determinant has the form
\begin{equation}\label{eq:a36}
{\det}_{n}(c,\Delta+\theta)=\prod_{r,s=1;1 \leq r s\leq
n}^{n}\left(\Delta+\theta-\Delta_{r,s}(c)\right)^{p(n-rs)} ,
\end{equation}
where $\Delta_{r,s}(c)$ is
\begin{equation}\label{eq:a37}
\Delta_{r,s}(c)=
\frac{1}{96}\left[(r+s)\sqrt{1-c}+(r-s)\sqrt{25-c}\right]^{2}
-\frac{1-c}{24} .
\end{equation}
and $p(n-rs)$ is the number of partitions of the integer $n-rs$.
In an ordinary CFT when the Kac determinant vanishes we have a
singular vector. In our case and for a rank 2 Jordan cell the
condition of vanishing $\det_{n}(c,\Delta+\theta)$ are:

(i) If $p(n-rs)\geq2$ for some $r$ and $s$, Kac determinant
vanishes for all values of $\Delta$ that satisfy in
$\Delta=\Delta_{r,s}(c)$.

(ii) If $p(n-rs)=1$ for some pairs of
$(r,s)=(r_{1},s_{1}),(r_{2},s_{2}),\cdots$ we can have vanishing
determinant if at least
$\Delta=\Delta_{r_{i},s_{i}}(c)=\Delta_{r_{j},s_{j}}(c)$. In this
case unlike (i) we are limited to special values for $\Delta$ and
$c$ which last condition is held. As an example we consider Kac
determinant at level 2. Since
\begin{equation}\label{eq:a38}
p(2-rs)=\cases{1&\mbox r=1 , s=1 \cr
               1&\mbox r=1 , s=2 or r=2 , s=1},
\end{equation}
all of them are cases of (ii). So
\begin{eqnarray}\label{eq:a39}
\Delta_{1,2}&=&\Delta_{2,1}\Rightarrow\left\{\begin{array}{rcl}
    c&=&1,\:\:\:\Delta=\frac{1}{4}\\
    c&=&25,\:\Delta=-\frac{5}{4},
    \end{array}\right. \nonumber\\
\Delta_{1,2}&=&\Delta_{1,1}\Rightarrow\:\:\:\:\: c=0,\Delta=0.
\end{eqnarray}
This approach can be extended easily to higher levels and Jordan
cells of bigger rank. These results are consistent with those of
\cite{floh3}.
\section{Four point functions}
To obtain further information about the theory with which we are
concerned, such as surface critical exponents, OPE structure,
monodromy group and etc. one should compute four point correlation
functions. In the language we have developed so far, the four
point correlation functions depend on four $\theta$'s in addition
to the coordinates of points
\begin{eqnarray}\label{eq:a40}
G(z_{1},z_{2},z_{3},z_{4},\theta_{1},\theta_{2},\theta_{3},
\theta_{4})&=&\left<\Phi_{1}(z_{1},\theta_{1})\ldots\Phi_{4}
(z_{4},\theta_{4})\right>\nonumber\\
&=&f(\eta,\theta_{1},\theta_{2},\theta_{3},\theta_{4})\prod_{1\leq
i<j\leq4}z_{ij}^{\mu_{ij}} .
\end{eqnarray}
where
\begin{eqnarray}\label{eq:a41}
\mu_{ij}&=&\frac{1}{3}\sum_{k=1}^{4}(\Delta_{k}+\theta_{k})
-(\Delta_{i}+\theta_{i})-(\Delta_{j}+\theta_{j}),
\:\:\:\:\:\:\:\:\:\:\:\:\:\:\:\:\:\:\:\:\:\:\:
\eta=\frac{z_{41}z_{23}}{z_{43}z_{21}}.
\end{eqnarray}

$\:\:\:\:\:\:\:\:\:\:\:\:\:\:\:\:\:\:$

This form is invariant under all conformal transformations.
Although there is no other restrictions on $G$ due to symmetry
considerations, but because of OPE structure, the four-point
function $\langle\phi\phi\phi\phi\rangle$ should vanish
\cite{mog,floh2}, that is, the term independent of $\theta_i$'s in
$G$ is zero. Thus in addition to the differential equations which
should be satisfied by $G$, one must impose the condition
$\langle\phi\phi\phi\phi\rangle=0$ on the solution derived.

If there is a singular vector in the theory, a differential
equation can be derived for
$f(\eta,\theta_{1},\theta_{2},\theta_{3},\theta_{4})$. Let us
consider a theory which contains a singular vector at level two.
As seen in previous section the singular vector in such a theory
is
\begin{equation}\label{eq:a42}
\chi^{(2)}(z_{4},\theta_{4})=\left[3
L_{-1}^{2}-(2(2\Delta_{4}+1)+4\theta_{4})L_{-2}\right]\Phi_{4}(z_{4},\theta_{4})
.
\end{equation}
As this vector is orthogonal to all the other operators in the
Verma module
\begin{eqnarray}\label{eq:a43}
\langle\Phi_{1}\Phi_{2}\Phi_{3}\chi^{(2)}\rangle&=&0 ,
\end{eqnarray}
one immediately is led to the differential equation

\begin{equation}\label{eq:a44}
\left[3\partial_{z_{4}}^{2}-(2(2\Delta_{4}+1)+4\theta_{4})
\sum_{i=1}^{3}\frac{\Delta_{i}+\theta_{i}}{(z_{i}-z_{4})^{2}}
-\frac{\partial_{z_{i}}}{z_{i}-z_{4}}\right]\langle\Phi_{1}
\Phi_{2}\Phi_{3}\Phi_{4}\rangle=0 .
\end{equation}
By sending points to
$z_{1}=0,\:\:\:z_{2}=1,\:\:\:z_{3}\longrightarrow\infty$ and
$z_{4}=\eta, $ we find
\begin{eqnarray}\label{eq:a45}
\partial_{\eta}^{2}f+[\frac{2\mu_{14}}{\eta}-\frac{2\mu_{24}}
{1-\eta}-\alpha\frac{2\eta-1}{\eta(1-\eta)}]\partial_{\eta}f
+[\frac{\mu_{14}(\mu_{14}-1)}{\eta^{2}}+\frac{\mu_{24}(\mu_{24}-1)}
{(1-\eta)^{2}}\nonumber\\
-\frac{2\mu_{14}\mu_{24}}{\eta(1-\eta)}-\frac{\alpha(\Delta_{1}+
\theta_{1}-\mu_{14})}{\eta^{2}}-\frac{\alpha(\Delta_{2}+\theta_{2}
-\mu_{24})}{(1-\eta)^{2}}+\frac{\alpha\mu_{12}}{\eta(1-\eta)}]f&=&0
,
\end{eqnarray}
where $\alpha=\frac{1}{3}\left[2(2\Delta_{4}+1)
+4\theta_{4}\right]$. Renormalizing using
\begin{eqnarray}\label{eq:a46}
H(\eta,\theta_{1},\theta_{2},\theta_{3},
\theta_{4})&=&\eta^{-\beta_{1}+\mu_{14}}(1-\eta)^{-\beta_{2}+\mu_{24}}f(\eta,\theta_{1},\theta_{2},\theta_{3},
\theta_{4}) ,
\end{eqnarray}
we find that $\beta_{i}$ satisfy
\begin{eqnarray}\label{eq:a47}
\beta_{i}(\beta_{i}-1)+\alpha(\beta_{i}-
\Delta_{i}-\theta_{i})&=&0 ,\:\:\:\:\:\:\:\:\:\:i=1,2
\end{eqnarray}
and $H$ satisfies the hypergeometric equation
\begin{eqnarray}\label{eq:a48}
\eta(1-\eta)\frac{d^{2}H}{d\eta^{2}}+[c-(a+b+1)\eta]\frac{d
H}{d\eta}- a b H&=&0,
\end{eqnarray}
where
\begin{eqnarray}\label{eq:a49}
ab&=&(\beta_{1}+\beta_{2})(\beta_{1}+
\beta_{2}+2\alpha-1)+\alpha(\Delta_{4}-
\Delta_{3}+\theta_{4}-\theta_{3}) ,\nonumber\\
a+b+1&=&2(\beta_{1}+\beta_{2}+\alpha) ,\nonumber\\
c&=&2\beta_{1}+\alpha .
\end{eqnarray}
We can now write down the solution of equation (48) in terms of
the hypergeometric series
\begin{equation}\label{eq:a50}
H(a,b,c;\eta)=K(\theta_{1},\theta_{2},\theta_{3},\theta_{4})h(a,b,c;\eta),
\end{equation}
where
\begin{equation}\label{eq:a51}
 h(a,b,c;\eta)=\sum_{n=0}^{\infty}\frac{(a)_{n}(b)_{n}}{n!(c)_{n}}\eta^{n} ,
\end{equation}
with $ (x)_{n}=x(x+1)\ldots(x+n-1)\:\:,(x)_{0}=1\:\:$. Also
\begin{equation}\label{eq:a52}
K(\theta_{1},\theta_{2},\theta_{3},\theta_{4})=
\sum_{i=1}^{4}k_{i}\theta_{i}+\sum_{1\leq
i<j\leq4}k_{ij}\theta_{i}\theta_{j}+ \sum_{1\leq
i<j<k\leq4}k_{ijk}\theta_{i}\theta_{j}\theta_{k}+k_{1234}
\theta_{1}\theta_{2}\theta_{3}\theta_{4}.
\end{equation}

Note that the coefficients in equation (51) contain nilpotent
terms. Therefore equation (51) actually describes more than one
solution. This expansion results in 16 functions. This is natural
because as seen in the case of two and three point functions,
inside the correlator (40) there exist sixteen distinct
correlation functions. Of course, if all the fields belong to the
same Jordan cell, only four of them may be independent and the
rest are related by crossing symmetry. Also, one of them which
only contains $\phi$ fields, vanishes and hence only three
independent functions remain. The form of these functions may be
obtained by expanding equation (50) and collecting powers of
$\theta_{i}$'s. Note that expanding equation (48) leads to
sixteen differential equations. The general form of these
equations is given in appendix. As an example we solve equation
(48) for the special case of $
\Delta_{1}=\Delta_{2}=\Delta_{3}=\Delta_{4}=\frac{1}{4}$. In this
case one of the solutions of equation (47) is
\begin{eqnarray}\label{eq:a53}
\beta_{1}&=&\frac{1}{2}+\theta_{1}-\frac{1}
{3}\theta_{4}+\frac{2}{3}\theta_{1}\theta_{4} ,\nonumber\\
\beta_{2}&=&\frac{1}{2}+\theta_{2}-\frac{1}{3}
\theta_{4}+\frac{2}{3}\theta_{2}\theta_{4} .
\end{eqnarray}
We then find from equation (49)
\begin{eqnarray}\label{eq:a54}
a&=&2+\theta_{1}+\theta_{2}+\theta_{3}+\theta_{4}
+\frac{2}{3}(\theta_{1}+\theta_{2}+\theta_{3}) \theta_{4} ,
\nonumber\\ b&=&1+\theta_{1}+\theta_{2}-\theta_{3}+
\frac{1}{3}\theta_{4}+\frac{2}{3}(\theta_{1}+\theta_{2}-\theta_{3})
\theta_{4} ,\nonumber\\
c&=&2+2\theta_{1}+\frac{2}{3}\theta_{4}+\frac{4}{3}\theta_{1}
\theta_{4} .
\end{eqnarray}
Finally from equations (40), (46) and (50) we get expressions for
the various four point functions. For example
\begin{eqnarray}\label{eq:a55}
\left<\psi(z_{1})\phi(z_{2})\phi(z_{3})\phi(z_{4})
\right>=k_{1}\eta^{\frac{2}{3}}(1-\eta)^{\frac{2}{3}}h_{0}(\eta)\prod_{1\leq
i<j\leq4}z_{ij}^{-\frac{1}{6}} .
\end{eqnarray}
Where $h_{0}(\eta)$ is given in the appendix. The other four point
functions having 2, 3 or 4 $\psi$'s can be calculated in the same
way. In these functions, depending on how many $\psi$'s are
present in the correlators, different functions appear on the
right hand side. If there is no $\psi$ in the correlator, the
correlator is zero, if there is one $\psi(z_{i})$, only
$H_{i}(=k_{i}h_{0}(\eta))$ appears, if there are two $\psi$'s,
$H_{i}$ and $H_{ij}$ appear, and so on. The situation is just the
same for the two or three point functions. For example the two
point function $\langle \phi (z) \psi(0) \rangle$ is written in
terms of the function $z^{-2\Delta}$ and in the correlator
$\langle \psi (z) \psi(0) \rangle$ there exist both functions
$z^{-2\Delta}$ and $z^{-2\Delta}$ln$z$.
\section{Energy momentum tensor}
Two central operators in a CFT are the energy momentum tensor,
$T$ with conformal weight $\Delta=2$ and the identity operator,
$I$ with conformal weight $\Delta=0$. However, $T$ is a secondary
field of the identity, because $L_{-2}I=T$.

 In an LCFT degenerate
operators exist which form a Jordan cell under conformal
transformation. This holds true for the identity as well. The
existence of a logarithmic identity operator has been discussed
by a number of authors \cite{gur,kogan,caux}.\\
Now consider the identity operator $\Omega$ and its logarithmic
partner $\omega$. According to equation (1) this pair transforms
as
\begin{eqnarray}\label{eq:a56}
\Omega(\lambda z)&=&\Omega(z) ,\nonumber\\ \omega(\lambda
z)&=&\omega(z)-\Omega(z)\ln \lambda .
\end{eqnarray}
So according to our convention, we define a primary field
$\Phi_{0}(z,\theta)$
\begin{equation}\label{eq:a57}
\Phi_{0}(z,\theta)=\Omega(z)+\theta \omega(z),
\end{equation}
with conformal weight $\theta$. Under scaling,
$\Phi_{0}(z,\theta)$ transforms according to equation
(\ref{eq:a3}). Thus we have
\begin{eqnarray}\label{eq:a58}
L_{0}\Omega(z)&=&0 ,\nonumber\\ L_{0}\omega(z)&=&\Omega(z),
\end{eqnarray}
where was first observed in $c=-2$ theory by Gurarie \cite{gur}.

Here we wish to find the field $T(z,\theta)$ with conformal weight
$2+\theta$ which is a secondary of $\Phi_{0}(z,\theta)$ in the
sense
\begin{equation}\label{eq:a59}
L_{-2}\Phi_{0}(z,\theta)=T(z,\theta) .
\end{equation}
By writing $T(z,\theta)=T_{0}(z)+\theta t(z)$ and since
$L_{0}T(z,\theta)=(2+\theta)T(z,\theta)$ we have
\begin{eqnarray}\label{eq:a60}
L_{0}T_{0}(z)&=&2T_{0}(z) ,\nonumber\\
L_{0}t(z)&=&2t(z)+T_{0}(z) .
\end{eqnarray}
This points to the existence of an extra energy momentum tensor
\cite{kogan,caux,lud}. By applying $L_{2}$ on both sides of
equation (59) we have
\begin{eqnarray}\label{eq:a61}
L_{2}T_{0}(z)&=&\frac{c}{2}\Omega(z) ,\nonumber\\
L_{2}t(z)&=&\frac{c}{2}\omega(z)+4\Omega(z) .
\end{eqnarray}
The first of this pair exists in an ordinary CFT, so $T_{0}(z)$
leads to the Virasoro algebra, while $t(z)$ must leads to a new
algebra \cite{lud}. We now attempt at finding the OPE of the
$T_{0}(z^{\prime})$ with $\psi(z)$ and extra energy momentum
tensor, $t(z)$. Because of OPE's invariance under scaling and
according to our convention it is sufficient to change conformal
weight of each field to $\Delta+\theta$. Consider the following
OPE
\begin{equation}\label{eq:a62}
T_{0}(z^{\prime})\Phi(z,\theta)=\frac{\Delta+\theta}{(z^{\prime}-z)^{2}}\Phi(z,\theta)+
\frac{\partial_{z}\Phi(z,\theta)}{z^{\prime}-z}+\cdots\:\:\:.
\end{equation}
This relation leads to the familiar OPE for
$T_{0}(z^{\prime})\phi(z)$ and a new OPE
\begin{equation}\label{eq:a63}
T_{0}(z^{\prime})\psi(z)=\frac{\phi(z)+\Delta\psi(z)}{(z^{\prime}-z)^{2}}+
\frac{\partial_{z}\psi(z)}{z^{\prime}-z}+\cdots\:\:\:.
\end{equation}
Also
\begin{equation}\label{eq:a64}
T_{0}(z^{\prime})T(z,\theta)=\frac{\frac{c(\theta)}{2}\Phi_{0}(z,\theta)}{(z^{\prime}-z)^{4}}
+\frac{2+\theta}{(z^{\prime}-z)^{2}}T(z,\theta)+\frac{\partial_{z}T(z,\theta)}{z^{\prime}-z}+\cdots\:\:\:,
\end{equation}
where $c(\theta)=c_{1}+\theta c_{2}$. Again we obtain two OPE, one
of them is $T_{0}(z^{\prime})T_{0}(z)$ which is known from CFT and
the other is
\begin{equation}\label{eq:a65}
T_{0}(z^{\prime})t(z)=\frac{\frac{c_1}{2}\omega(z)+\frac{c_2}{2}\Omega(z)}{(z^{\prime}-z)^{4}}
+\frac{T_{0}(z)+2t(z)}{(z^{\prime}-z)^{2}}+\frac{\partial_{z}t(z)}{z^{\prime}-z}+\cdots\:\:\:.
\end{equation}

The emergence of an extra energy momentum tensor and central
charge have been noticed by Gurarie and Ludwig \cite{lud},
although our approach is very different. Recently it has been
argued \cite{nic} that in theories with zero central charge
($c=0$) $t$, the logarithmic partner of $T_{0}$ is not a
descendant of any other field. So these theories can have non
degenerate vacua. Such theories will not fit into the framework
presented here. Theories with non zero central charge ($c\ne0$)
behave very differently. In these theories there exist a
logarithmic partner for $T_{0}$ only if the vacua is degenerate
and so $t$ is a descendant field \cite{nic}.

It is worth noting that equations (56) imply that $\langle \Omega
\rangle$ vanishes whereas $\langle\omega\rangle = 1$ . This
immediately results in the vanishing of $\langle T_{0}T_{0}
\rangle $, even though the central charge may not vanish.

\section{Boundary}
Let us now consider the problem of LCFT near a boundary. As shown
in \cite{cardy} in an ordinary CFT, if the real axis is taken to
be the boundary, with certain boundary condition that
$T=\overline{T}$ on the real axis, the differential equation
satisfied by n-point function near a boundary are the same as the
differential equations satisfied by 2n-point function in the
bulk. This trick may be used in order to derive correlations of
an LCFT near a boundary \cite{kog,MR}. Here we rederive the same
results using the nilpotent formalism. Again we consider an LCFT
with a rank 2 Jordan cell. First we find the one point functions
of this theory. By applying $L_{0},L_{\pm1}$ on the correlators,
one obtains
\begin{eqnarray}\label{eq:a66}
(\partial_{z}+\partial_{\bar{z}})\langle\Phi(z,\bar{z},\theta)
\rangle&=&0 ,\nonumber\\
(z\partial_{z}+\bar{z}\partial_{\bar{z}}+2(\Delta+\theta)
\langle\Phi(z,\bar{z},\theta\rangle&=&0 ,\nonumber\\
(z^{2}\partial_{z}+\bar{z}^{2}\partial_{\bar{z}}+
2z(\Delta+\theta)+2\bar{z}(\Delta+\theta))\langle
\Phi(z,\overline{z},\theta\rangle&=&0 .
\end{eqnarray}
In these equations, we have assumed that $\Phi$ is a scalar field
so that $\Delta=\bar{\Delta}$. The first equation states
$\langle\Phi(z,\bar{z},\theta)\rangle$ is a function of
$z-\bar{z}$ and the solution to the second equation is
\begin{eqnarray}\label{eq:a67}
\langle\Phi(y,\theta)\rangle&=&\frac{f(\theta)}{y^{2(\Delta+\theta)}}
,
\end{eqnarray}
where $y=z-\bar{z}$. The third line of equation (66) is
automatically satisfied by this solution. Expanding $f(\theta)$
as $a+b\theta$ one finds
\begin{eqnarray}\label{eq:a68}
\langle\Phi(y,\theta)\rangle&=&\frac{a}{y^{2\Delta}}+\frac{\theta}{y^{2\Delta}}(b-2a\ln
y).
\end{eqnarray}
As the field $\Phi(y,\theta)$ is decomposed to
$\phi(y)+\theta\psi(y)$ one can read the one-point functions
$\langle\phi(y)\rangle$ and $\langle\psi(y)\rangle$ from the
equation (68)
\begin{eqnarray}\label{eq:a69}
\langle\phi(y)\rangle&=&\frac{a}{y^{2\Delta}},\nonumber\\
\langle\psi(y)\rangle&=&\frac{1}{y^{2\Delta}}(b-2a\ln y) .
\end{eqnarray}
To go further, one can investigate the two-point function
\begin{equation}
G(z_{1},\bar{z}_{1},z_{2},\bar{z}_{2},\theta_{1},\theta_{2})=
\langle\Phi(z_{1},\bar{z}_{1},\theta_{1})\Phi(z_{2},
\bar{z}_{2},\theta_{2})\rangle,
\end{equation}
 in the same theory. Invariance
under the action of $L_{-1}$ implies
\begin{eqnarray}\label{eq:a71}
(\partial_{z_{1}}+\partial_{\bar{z}_{1}}+\partial_{z_{2}}+\partial_{\bar{z}_{2}})G&=&0
.
\end{eqnarray}
The most general solution of this equation is
$G=G(y_{1},y_{2},x_{1},x_{2},\theta_{1},\theta_{2})$, where
$y_{1}=z_{1}-\bar{z}_{1},\:\:\:\:y_{2}=z_{2}-\bar{z}_{2},
\:\:\:\:x=x_{2}-x_{1}$ and $x_{i}=z_{i}+\bar{z}_{i}$. By
invariance under the action of $L_{_{0}}$ we should have
\begin{eqnarray}\label{eq:a72}
\left[y_{1}\frac{\partial}{\partial
y_{1}}+y_{2}\frac{\partial}{\partial
y_{2}}+x\frac{\partial}{\partial
x}+2(\Delta+\theta_{1})+2(\Delta+\theta_{2})\right]G&=&0 ,
\end{eqnarray}
which implies
\begin{eqnarray}\label{eq:a73}
G&=&\frac{1}{x^{4\Delta+2\theta_{1}+2\theta_{2}}}
f(\alpha_{1},\alpha_{2},\theta_{1},\theta_{2}),
\end{eqnarray}
where $\alpha_{1}=\frac{y_{1}}{x}$ and
$\alpha_{2}=\frac{y_{2}}{x}$. Now consider the action of $L_{1}$
on $G$
\begin{eqnarray}\label{eq:a74}
(x_{1}+x_{2})\left[y_{1}\frac{\partial}{\partial
y_{1}}+y_{2}\frac{\partial}{\partial
y_{2}}+x\frac{\partial}{\partial
x}+2(\Delta+\theta_{1})+2(\Delta+\theta_{2})\right]G\nonumber\\
+\left[xy_{1}\frac{\partial}{\partial
y_{1}}-xy_{2}\frac{\partial}{\partial
y_{2}}+(y_{1}^{2}-y_{2}^{2})\frac{\partial}{\partial x}
+2x(\theta_{1}-\theta_{2})\right]G=0 .
\end{eqnarray}
The first bracket is zero because of equation (72). Substituting
the solution (73) in equation (74) the function $f$ satisfies
\begin{eqnarray}\label{eq:a75}
\left(\alpha_{1}+\frac{\alpha_{1}}{\alpha_{1}^{2}-
\alpha_{2}^{2}}\right)\frac{\partial f }{\partial
\alpha_{1}}+\left(\alpha_{2}+\frac{\alpha_{2}}{\alpha_{2}^{2}
-\alpha_{1}^{2}}\right)\frac{\partial f }{\partial
\alpha_{2}}+2\left(2\Delta+\theta_{1}+\theta_{2}+\frac{\theta_{1}
-\theta_{2}}{\alpha_{1}^{2}-\alpha_{2}^{2}}\right)f=0.
\end{eqnarray}
The most general solution of above is
\begin{eqnarray}\label{eq:a76}
f(\alpha_{1},\alpha_{2},\theta_{1},\theta_{2})&=&\frac{1}{(\alpha_{1}\alpha_{2})^
{2\Delta+\theta_{1}+\theta_{2}}}
\left(\frac{\alpha_{2}}{\alpha_{1}}\right)^{\theta_{1}-\theta_{2}}
g\left(\frac{1+\alpha_{1}^{2}+\alpha_{2}^{2}}{\alpha_{1}
\alpha_{2}},\theta_{1},\theta_{2}\right),
\end{eqnarray}
where $g$ is an arbitrary function. So the two point function $G$
is found up to an unknown function
\begin{eqnarray}\label{eq:a77}
\langle \Phi(z_{1},\bar{z}_{1},\theta_{1})\Phi(z_{2},\bar{z}_{2},
\theta_{2})\rangle&=&
\frac{1}{(y_{1}y_{2})^{2\Delta+\theta_{1}+\theta_{2}}}
\left(\frac{y_{1}}{y_{2}}\right)^{\theta_{1}-\theta_{2}}
g\left(\frac{x^{2}+y_{1}^{2}+y_{2}^{2}}{y_{1}y_{2}},\theta_{1},\theta_{2}\right),
\end{eqnarray}
which is the same as the solution obtained in \cite{MR}.
\section{Extension of nilpotent variable to a four component superfield}
In section 1 we introduced a two component superfield $
\Phi(z,\theta)=\phi(z)+\theta\psi(z)$ which its components forms
a Jordan cell under conformal transformations. In this section we
extend our superfield to a four component one, exploiting a
grassman variable $\eta$
\begin{equation}\label{eq:a88}
\Phi(z,\eta)= \phi(z) + \bar{\alpha}(z) \eta +\bar{\eta}
\alpha(z) + \bar{\eta} \eta\psi(z).
\end{equation}
Here a fermionic field $\alpha(z)$ with the same conformal
dimension as $\phi(z)$ has been added to the multiplet, and the
nilpotent variable $\theta$ has been interpreted as
$\bar{\eta}\eta$ . Note that both $\alpha$ and $\bar{\alpha}$
live in the holomorphic section of the theory. Now we observe
that $ \Phi (z,\eta )$ has the following transformation law under
scaling
\begin{eqnarray}\label{eq:a89}
\Phi(\lambda z,\eta)&=&\lambda^{-(\Delta+\bar{\eta}\eta)}
\Phi(z,\eta) .
\end{eqnarray}
To find out what this scaling law means, one should expand both
sides of equation (\ref{eq:a89}) in terms of $\eta$ and
$\bar{\eta}$. Doing this and comparing the two sides of equation
(\ref{eq:a89}), it is found that $\phi(z)$ and $\psi(z)$
transform as equation (\ref{eq:a1}) and $\alpha$ and
$\bar{\alpha}$ are ordinary fields of dimension $\Delta$. The
appearance of such fields has been proposed by Kausch
\cite{kausch}, within the $c=-2$ theory.

In previous sections we saw that using this structure one can
derive most of the properties of LCFTs. Let us first consider the
identity operator. In a two component superfield, in addition to
the ordinary unit operator $\Omega$, with the property $\Omega S
= S$ for any field $S$, and its logarithmic partner, which we
denoted by $\omega$, there exist  two other fields with zero
conformal dimension for the multiplet to complete
\begin{equation}\label{eq:a90}
\Phi_0(z,\eta ) = \Omega  + \bar{\xi}(z) \eta +\bar{\eta}
\xi(z)+\bar{\eta} \eta \omega(z),
\end{equation}
with the property that $\langle\Phi_0(z,\eta)\rangle =\bar{\eta}
\eta$. Note that LCFTs have the curious property that $\langle
\Omega\rangle = 0$. The existence of these fields and their OPEs
have been discussed by Kausch \cite{kausch}. Kausch takes the
ghost action of

\begin{equation}\label{eq:a91}
S=\frac{1}{\pi}\int  d^{2}z
(a\bar{\partial}b+\bar{a}\partial\bar{b}),
\end{equation}
with $c=-2$. Instead of the degrees of freedom $a$ and $b$, he
takes two fields $\chi^{\alpha}$ on equal footing, where
$\alpha=1,2$. Taking the Laurent expansion
\begin{equation}\label{eq:a92}
\chi^\alpha =\sum_n \chi^{\alpha}_{n} z^{-n-1},
\end{equation}
he observes that $\chi^{\alpha}_{0}$ plays the role of a ladder
operator for the multiplet defined in equation (\ref{eq:a90})
\begin{eqnarray}\label{eq:a93}
\chi^{\alpha}_{0}\Omega=0, \nonumber\\
\chi^{\alpha}_{0}\omega=\xi^{\alpha},\nonumber\\
\chi^{\alpha}_{0}\xi^{\beta}=d^{\alpha\beta}\Omega,
\end{eqnarray}
 where $d^{\alpha\beta}$ is a totally antisymmetric matrix. Note
 that over the vacuum multiplet we have
\begin{equation}\label{eq:a96}
\chi^{\alpha}_{0}\chi^{\beta}_{0}=d^{\alpha\beta}L_0.\hspace{2mm}
\end{equation}

An interesting question is whether it is possible to write the
OPEs of the fields in the multiplet in terms of our fields $\Phi_0
(z,\eta)$. The OPE of $\Phi_0$ with itself has to give back
$\Phi_0$ to lowest order

\begin{equation}\label{eq:a97}
{\Phi}_{0} (z_1,\eta_1) \Phi_0(z_2,\eta_2) \sim
(z_1-z_2)^{\bar{\eta}_1 \eta_2 +\bar{\eta}_2 \eta_1}
\Phi_0(z_1,\eta_3),
\end{equation}
where $\eta_3= \eta_1 + \eta_2$. To see why this OPE has been
proposed, one can look at the behaviour of both sides of equation
(\ref{eq:a97}) under scaling transformation. Under the
transformation $z \rightarrow \lambda z$ the LHS of equation
(\ref{eq:a97}) transforms as $LHS \rightarrow
\lambda^{-(\bar{\eta}_1 \eta_1 + \bar{\eta}_2 \eta_2)} LHS$, (See
the transformation law (\ref{eq:a89})) and the RHS of equation
(\ref{eq:a97}) transforms as $RHS \rightarrow
\lambda^{\bar{\eta}_1 \eta_2 +\bar{\eta}_2 \eta_1- \bar{\eta}_3
\eta_3} RHS$, which are the same. So the OPE proposed here seems
reasonable.

The OPEs of the primary fields involved can then be extracted by
expanding in powers of $\eta$. Taking $z_2=0$ and renaming
$z_1=z$ the OPEs of these fields, aside from trivial OPEs of
$\Omega$, are
\begin{eqnarray}\label{eq:a98}
\bar{\xi}(z)\xi(0)&=&2i(\omega+\Omega\log{z}),\nonumber\\
\xi(z)\omega(0)&=&-\xi \log{z},\nonumber\\
\omega(z)\omega(0)&=&-\log{z} (2\omega+\Omega \log{z}).
\end{eqnarray}
These results are consistent with those of \cite{kausch}. An
obvious generalization of (\ref{eq:a97}) leads to
\begin{equation}\label{eq:a101}
\Phi_{a}(z_1,\eta_1) \Phi_{b}(z_2,\eta_2) \sim
(z_1-z_2)^{\Delta_{c}- \Delta_{b} - \Delta_{a} + \bar{\eta}_1
\eta_2 + \bar{\eta}_2 \eta_1} C_{abc} \Phi_{c}(z_1,\eta_3).
\end{equation}
for any three primary fields with arbitrary conformal dimensions.
An immediate implication of these equations is that the OPE of two
$\phi$ fields will be the first components of the multiplets such
as an $\Omega$ or another $\phi$, thus the expectation values of
an arbitrary string $\langle
\phi_{1}\phi_{2}\cdots\phi_{n}\rangle$ will vanish in any LCFT
\cite{floh2}.

A consequence of existence of $\Phi_0$ is the existence of the
energy momentum multiplet
\begin{equation}\label{eq:a102}
T(z,\eta)=L_{-2} \Phi_0(z,\eta).
\end{equation}
We thus observe that we have three partners for the energy
momentum tensor $T(z,\eta)=T_0(z)+\bar{\eta}
\zeta(z)+\bar{\zeta}(z)\eta + \bar{\eta}\eta t(z)$, as discussed
by Gurarie and Ludwig \cite{lud}. In their paper, they have
considered some specific theories and have suggested some OPEs for
different fields of energy momentum tensor multiplet. However,
getting some insight from the OPEs we have written so far, we
propose
\begin{eqnarray}\label{eq:a103}
T(z,\eta_1)T(0,\eta_2) = \hspace{100mm}\nonumber \\
 z^{ \bar{\eta}_1 \eta_2 + \bar{\eta}_2
\eta_1} \left[ \frac{\frac{c(\eta_3)}{2}\Phi_0(\eta_3)}{z^{4}} +
\frac{d(\eta_3)\chi(\eta_3)}{z^3}+
\frac{e(\eta_3)T(\eta_3)}{z^{2}} + \frac{f(\eta_3)
\partial_z{T(\eta_3)}}{z}\right].
\end{eqnarray}
There are some points in this OPE which should be clarified.
First of all, in contrast with the OPE of ordinary energy momentum
tensor, there exists a $1/z^3$ term. In the ordinary OPE, this
term vanishes because $L_{-1}\Omega=0$. Such a reasoning can not
be extended to the case of other fields of the energy momentum
multiplet, that is, one can not assume that $L_{-1}\omega$,
$L_{-1}\xi$ and $L_{-1}\bar{\xi}$ all vanish. In equation
(\ref{eq:a103}) we have denoted $L_{-1}\Phi_0(\eta)$ by
$\chi(\eta)=\bar{\eta}\sigma + \bar{\sigma}\eta + \bar{\eta}\eta
J $.

 The other point is that the constants appearing on the RHS
depend only on $\eta_3$. This seems a reasonable assumption. The
constants such as $c(\eta)$ have only two components, that is we
can express them as $c(\eta)=c_1+c_2 \bar{\eta} \eta$, since we
wish to avoid non scalar constants in our theory. Note that $c_1$
corresponds to the usual central charge in the ordinary theories.
It will be more clear when one looks at the OPE of two $T_0$'s
which is given below. The constant $e(\eta)$ is taken to be
$e(\eta)=2+\bar{\eta} \eta$ for consistency with known conformal
dimension of $T_0(z)$ and with the fact that $t(z)$ is the
logarithmic partner of $T_0(z)$. Also $f(\eta)$ is taken to be
unit, in order to obtain the familiar action of $T_{0}$ on the
members of the multiplet. By now, there is no restrictions on
$d(\eta)$ and we will take it to be $d_1+\bar{\eta}\eta d_2$,
however as shown below, the constant $d_2$ plays no role in the
OPE.

Expanding both sides of equation (\ref{eq:a103}), we find the
explicit form of the OPEs
\begin{eqnarray}\label{eq:a104}
T_0(z)T_0(0)&=& \frac{ \frac{c_1}{2} \Omega}{z^{4}}+
\frac{2T_0}{z^{2}}+ \frac{\partial_z T_0}{z}, \nonumber \\
T_0(z)t(0)&=&\frac{ \frac{c_2}{2} \Omega+
\frac{c_1}{2}\omega}{z^{4}} + \frac{d_1 J}{z^3} + \frac{2t +
T_0}{z^{2}} + \frac{
\partial_z t}{z},\nonumber\\
t(z)t(0)&=&-\frac{1}{2} \frac{ \log{z}\left((c_1 \log{z} +
2c_2)\Omega + 2c_1 \omega\right)}{z^{4}} -\frac{2d_1 \log{z}
J}{z^3}
\nonumber \\
&-&\frac{ 2 \log{z}(2t +T_0 \log{z}) + 2 \log{z}
T_0}{z^{2}}-\frac{
\partial_z( \log{z}
(2t + T_0 \log{z}))}{z},\nonumber \\
T_0(z)\zeta(0)&=&\frac{1}{2} \frac{c_1 \xi}{z^{4}} + \frac{d_1
\sigma}{z^3}+\frac{2
\zeta}{z^{2}}+\frac{\partial_z \zeta}{z},\nonumber\\
\zeta(z)t(0)&=&-\frac{1}{2} \frac{c_1 \xi \log{z}}{ z^{4}}
-\frac{d_1 \log{z} \sigma}{z^3}- \frac{2\zeta\log{z}}{z^{2}}-
\frac{\partial_z (\zeta \log{z})}{z},\nonumber\\
 \bar{\zeta}(z) \zeta(0)&=&-\frac{1}{2} \frac{ c_1 (\omega +
\Omega \log{z}) + c_2\Omega}{z^{4}}-\frac{d_1 J}{z^3}-\frac{2(t +
T_0 \log{z}) + T_0}{z^{2}}\nonumber\\
&-&\frac{\partial_z( t + T_0 \log{z})}{z}.
\end{eqnarray}
We observe that these expressions are not the same as those found
by \cite{lud}. Aside from the presence of the $1/z^3$ terms, the
most important difference is that we have the logarithmic partner
of unity operator in our OPEs as well as unity, itself. The
existence of 'pseudo-unity' when there is a logarithmic partner
for the energy momentum tensor is obligatory. In fact one can
easily show that $L_{2}t(z)$ is the logarithmic partner of unity.
The presence of pseudo-unity causes some problems with the OPEs
derived by \cite{lud}, because now the expectation value of
$\Omega$ is zero. This leads to

\begin{equation}\label{eq:a110}
\langle T_0(z_1)T_0(z_2) \rangle =0
\end{equation}
regardless whether $c_1$ is zero or not. Instead we have
\begin{equation}\label{eq:a111}
\langle T_0(z_1)t(z_2)\rangle = \frac{c_1}{2(z_1-z_2)^4}
\end{equation}
which is just the one written by \cite{lud} with $b$ being $c_1/2$
of our theory. So it is observed that in such theories, there is
no need to set $c_1=0$. However as suggested by \cite{lud}, two
central charges appear as is seen in equation (\ref{eq:a103}).
Since this structure is reminiscent of supersymmetry, most
authors have set $c_1=0$. Setting $c_1$ to zero, we do find a
much simpler OPE. However, two obvious differences with
supersymmetry may be observed. First, $\zeta$ does not transform
like the supercurrent of SCFT. Second, the role of $t(z)$ is not
clear. We therefore believe that connection with supersymmetry,
if any, has to appear at a deeper level. Thus identifying $t$
with the similar object in \cite{lud} may not be right.

In this section we saw that the idea of grassman variables in
LCFT can be extended to include fermionic fields in the theory,
and this naturally leads to a current algebra involving the energy
momentum tensor, its logarithmic partner and two fermionic
currents. Despite the superficial resemblance to supersymmetry
such as super multiplets etc., there is need of further
clarification if there is any supersymmetry within the theory.
\section{AdS/LCFT correspondence and correlation functions}
Much work has been done in the last few years based on the AdS/CFT
correspondence with the aim of understanding conformal field
theories \cite{aha}. Within this framework, the correlation
functions of operators on the boundary of Anti de Sitter space
are determined in terms of appropriate bulk propagators. While
the form of the two and three point functions within CFT are
fixed by conformal invariance, it is interesting to find actions
in the bulk which result in the desired boundary green functions.
In particular it is interesting to discover which actions give
rise to logarithmic conformal field theories which leads to
AdS/LCFT correspondence. Two points should be clarified, first
what is meant by ordinary AdS/CFT correspondence and second what
is an LCFT, and how does it fit into the correspondence.

The conjecture states that a correspondence between theories
defined on AdS$_{d+1}$ and CFT$_{d}$ can be found. Suppose that a
classical theory is defined on the AdS$_{d+1}$ via the action
$S[\Phi]$. On the boundary of this space the field is constrained
to take certain boundary value ${\Phi|}_{\partial AdS} =\Phi_0$.
With this constraint, one can calculate the partition function
\begin{eqnarray}\label{eq:a112}
Z[\Phi_0]={e^{-S_{Cl}[\Phi]}}|_{\Phi_{\partial AdS}=\Phi_0}.
\end{eqnarray}
On the other hand in the CFT$_{d}$ space, there exist operators
like $O$ which belong to some conformal tower. Now the
correspondence states that the partition function calculated in
AdS is the generating function of the theory in CFT with $\Phi_0$
being the source, that is

\begin{equation}\label{eq:a113}
Z[\Phi_0]=\left\langle e^{\int O\Phi_0} \right\rangle.
\end{equation}

How do logarithmic conformal field theories fit into this
picture? The bulk actions defined on AdS$_{3}$ which give rise to
logarithmic operators on the boundary where first discussed in
\cite{khor,iko} and have consequently been discussed by a number
of authors \cite{lewis,myu}. More recently a connection with world
sheet supersymmetry has been discussed in \cite{pol}. In this
section the correspondence is explained explicitly and the two
point correlation functions of different fields of CFT are
derived \cite{rou}.

 To begin, one should propose an action on AdS. As we will have an
operator like
\begin{equation} \label{eq:a115}
O(z,\eta)= A(z) +\bar{\zeta}(z) \eta +\bar{\eta} \zeta(z) +
\bar{\eta} \eta B(z),
\end{equation}
with scaling law
\begin{eqnarray}\label{eq:a116}
O(\lambda z,\eta)=\lambda^{-(\Delta+\bar{\eta}\eta)} O(z,\eta) .
\end{eqnarray}
in LCFT part of the theory, there should be a corresponding field
in AdS, $\Phi(x,\eta)$, which can be expanded as

\begin{equation}\label{eq:a117}
\Phi(x,\eta)=C(x)+\bar{\eta}\beta(x)+\bar{\beta}(x)\eta+
\bar{\eta}\eta D(x),
\end{equation}
where $x$ is $(d+1)$ dimensional with components $x^0,\cdots,x^d$.
Of course $d=2$ is the case we are most interested in, any how the
result can be applied to any dimension and so we will consider
the general case in our calculations. Let us then consider the
action
\begin{equation}\label{eq:a118}
S=-\frac 12\int d^{d+1}x\int d\bar{\eta} d\eta [(\nabla \Phi
(x,\eta )).(\nabla \Phi (x,-\eta ))+ m^2(\eta )\Phi(x,\eta )\Phi
(x,-\eta )].
\end{equation}
This action seems to be the simplest non trivial action for the
field $\Phi (x,\eta ).$ To write it explicitly in terms of the
four components of the the field, one should expand equation
(\ref{eq:a118}) in powers of $\bar{\eta}$ and $\eta$ using
equation (\ref{eq:a117}). Integrating over $\bar{\eta}$ and $\eta$
one finds
\begin{equation}  \label{eq:a119}
S=-\frac{1}{2}\int d^{d+1} x [2(\nabla C).(\nabla D)+ 2{m^2}_1CD
+{m^2}_2 C^2 + 2(\nabla\bar{\beta}).(\nabla\beta)+2{m^2}_1
\bar{\beta}\beta].
\end{equation}
To derive expression (\ref{eq:a119}) we have assumed
$m^2(\eta)={m^2}_1 +{m^2}_2 \bar{\eta} \eta$. Note that the
bosonic part of this action is the same as the one proposed by
\cite{khor,iko} with ${m^2}_1=\Delta(\Delta-d)$ and ${
m^2}_2=2\Delta-d$. In our theory with proper scaling of the
fields, one can recover these relations. The equation of motion
for the field $\Phi$ is
\begin{equation}  \label{eq:a120}
({\nabla}^2 - m^2(\eta))\Phi(x,\eta)=0.
\end{equation}
The Dirichlet Green function for this system satisfies the
equation
\begin{equation}  \label{eq:a121}
({\nabla}^2 - m^2(\eta))G(x,y,\eta)=\delta(x,y),
\end{equation}
together with the boundary condition
\begin{equation}  \label{eq:a122}
G(x,y,\eta)|_{x\in \:\partial_{AdS}} =0.
\end{equation}
With this Green function, the Dirichlet problem for $\Phi$ in AdS
can be solved readily. However, near the boundary of AdS, that is
$x^d \simeq 0$,  the metric diverges so the problem should be
studied more carefully. One can first solve the problem for the
boundary at $x^d=\varepsilon$ and then let $\varepsilon$ tend to
zero. With properly redefined scaled fields at the boundary one
can avoid the singularities in the theory. So we first take the
boundary at $ x^d=\varepsilon$ and find the Green function
\cite{muck}
\begin{equation}\label{eq:a123}
G(x,y,\eta)|_{y^d=\varepsilon}=-a(\eta)\varepsilon^
{\Delta+\bar{\eta}\eta-d}
\left(\frac{x^d}{(x^d)^2+|\mathbf{x-y}|^2}\right)^
{\Delta+\bar{\eta}\eta},
\end{equation}
where the bold face letters are $d$-dimensional and live on the
boundary. The field in the bulk is related to the boundary fields
by
\begin{equation}  \label{eq:a124}
\Phi(x,\eta)=2a(\eta)(\Delta+\bar{\eta}\eta)\varepsilon^{\Delta+\bar{\eta}
\eta-d} \int_{y^d=\varepsilon} d^d y\:
\Phi(\mathbf{y},\varepsilon,\eta)
\left(\frac{x^d}{(x^d)^2+|\mathbf{x-y}|^2}\right)^
{\Delta+\bar{\eta}\eta},
\end{equation}
with $a=\frac{\Gamma(\Delta+\bar{\eta}\eta)}{2\pi^{d/2}
\Gamma(\alpha+1)}$ and $\alpha=\Delta+\bar{\eta}\eta-d/2$. To
compute Gamma functions in whose argument appears
$\bar{\eta}\eta$, one should make a Taylor expansion for the
function, that is
\begin{equation}\label{eq:a125}
\Gamma(a+\bar{\eta}\eta)=\Gamma(a)+\bar{\eta}\eta\Gamma^{\prime}(a).
\end{equation} There is no higher terms in this Taylor expansion
because $(\bar{\eta} \eta)^2=0$. Now defining
$\Phi_b(\mathbf{x},\eta)=\lim_{\varepsilon\rightarrow0} (\Delta+
\bar{\eta}\eta)\varepsilon^{\Delta+\bar{\eta}\eta-d}\Phi(\mathbf{x}
,\varepsilon,\eta)$, we have
\begin{equation}\label{eq:a126}
\Phi(x,\eta)=\int d^d \mathbf{y}
\left(\frac{x^d}{(x^d)^2+|\mathbf{x-y}|^2} \right)^
{\Delta+\bar{\eta}\eta} \Phi_b(\mathbf{y},\eta).
\end{equation}
Using the solution derived, one should compute the classical
action. First note that the action can be written as (using the
equation of motion and integrating by parts)
\begin{equation}  \label{eq:a127}
S_{cl.}=\frac{1}{2} \lim_{\varepsilon\rightarrow 0}
\varepsilon^{1-d} \int d \bar{\eta}d \eta \int d^d \mathbf{y}
\left[\Phi(\mathbf{y},\varepsilon,\eta) \frac{\partial
\Phi(\mathbf{y},\varepsilon,-\eta)}{\partial x^d}\right],
\end{equation}
putting the solution (\ref{eq:a126}) into equation
(\ref{eq:a127}), the classical action becomes
\begin{equation}  \label{eq:a128}
S_{cl.}(\Phi_b)=\frac{1}{2}\int d \bar{\eta}d \eta \int d^d
\mathbf{x} d^d \mathbf{y} \frac{a(\eta)
\Phi_b(\mathbf{x},\eta)\Phi_b(\mathbf{y},-\eta)}{|
\mathbf{x-y}|^{2\Delta+2\bar{\eta}\eta}}.
\end{equation}

The next step is to derive correlation functions of the operator
fields on the boundary by using AdS/CFT correspondence. In our
language the operator $O$ has an $\eta $ dependence, in addition
to its usual coordinate dependence. It lives in the LCFT space
and can be expanded as
\begin{equation}\label{eq:a129}
O(\mathbf{x},\eta)=A(\mathbf{x})+\bar{\eta}\zeta(\mathbf{x})+
\bar{\zeta}(\mathbf{x})\eta +\bar{\eta}\eta B(\mathbf{x}),
\end{equation}
so the AdS/LCFT correspondence becomes
\begin{equation}\label{eq:a130}
\left\langle \exp\left( \int d\bar{\eta}d\eta \int
d^{d}\mathbf{x}O(\mathbf{x},\eta )\Phi _{b} (\mathbf{x},\eta
)\right) \right\rangle =e^{S_{cl}(\Phi _{b})}. \label{AdS/LCFT}
\end{equation}
Expanding both sides of  this equation in powers of $\Phi _{b}$
and integrating over $\eta $'s, the two point function of
different components of $O(\mathbf{x},\eta )$ can be found
\begin{eqnarray}\label{eq:a131}
\langle A(\mathbf{x})A(\mathbf{y})\rangle &=&0,\nonumber\\
\langle A(\mathbf{x})B(\mathbf{y})
\rangle &=&\frac{a_{1}} {(\mathbf{x}-\mathbf{y})^{2\Delta }},\nonumber\\
\langle B(\mathbf{x})B(\mathbf{y})\rangle  &=&
\frac{1}{(\mathbf{x}-\mathbf{y})^{2\Delta }}(a_{2}-2a_{1}\log
(\mathbf{x}-\mathbf{y})),\nonumber\\
\langle \bar{\zeta}(\mathbf{x})\zeta(\mathbf{y})\rangle
&=&\frac{-a_{1}}{(\mathbf{x}-\mathbf{y})^{2\Delta }},
\end{eqnarray}
with all other correlation functions being zero. These
correlation functions can be obtained in another way. Knowing the
behaviour of the fields under conformal transformations, the form
of two-point functions are determined. The scaling law is given
by equation (\ref{eq:a116}). Using this scaling law, most of the
correlation functions derived here are fulfilled. However it does
not lead to vanishing correlation functions of $ \langle
A(\mathbf{x})\zeta(\mathbf{y})\rangle $ and $\langle
B(\mathbf{x})\zeta(\mathbf{y}) \rangle $. These correlation
functions are found to be
\begin{eqnarray}\label{eq:a135}
 \langle
A(\mathbf{x})\zeta(\mathbf{y})\rangle &=& \frac{b_1}
{\mathbf{(x-y)}^{2\Delta}},\nonumber
\\
\langle B(\mathbf{x})\zeta(\mathbf{y})\rangle &=&
\frac{1}{\mathbf{(x-y)}^{2\Delta}} \left( b_2 -2b_1
\log\mathbf{(x-y)} \right).
\end{eqnarray}

Of course, assuming $b_1=b_2=0$ one finds the forms derived above.
However, the vanishing value of such correlators comes from some
other properties of the theory. What forces these constants to
vanish is the fact that the total fermion number is odd
\footnote{this observation is due to M. Flohr}. One way of seeing
this is to look at the OPE as given in \cite{mog}. The OPE of two
$O$-fields has been proposed to be
\begin{equation}\label{eq:a137}
O(z)O(0)\sim z^{\bar{\eta_1}\eta_2+\bar{\eta_2}\eta_1}
\frac{\Phi_0(\eta_3)}{z^{2(\Delta+\bar{\eta_3}\eta_3)}},
\end{equation}
where $\eta_3=\eta_1+\eta_2$ and $\Phi_0$ is the identity
multiplet
\begin{equation}\label{eq:a138}
\Phi_0(\eta)=\Omega +
\bar{\eta}\xi+\bar{\xi}\eta+\bar{\eta}\eta\omega,
\end{equation}
with the property
\begin{equation}\label{eq:a139}
\langle\Phi_0(\eta)\rangle=\bar{\eta}\eta.
\end{equation}
Note that the ordinary identity operator is $\Omega$ which has
the unusual property that $\langle\Omega\rangle=0$, but its
logarithmic partner, $\omega$, has nonvanishing norm. Calculating
the expectational value of both sides of equation
(\ref{eq:a137}), the correlation functions of different fields
inside $O$ are found which leads to vanishing correlators $
\langle A(\mathbf{x})\zeta(\mathbf{y})\rangle $ and $\langle
B(\mathbf{x})\zeta(\mathbf{y}) \rangle $, just the same result as
derived by ADS/LCFT correspondence.
\section{BRST symmetry of the theory}
The existence of some ghost fields in the action of the theory
considered so far, is reminiscent of BRST symmetry. A few remarks
clarifying the word "ghost" may be in order here. The action
defined by equation (\ref{eq:a119}) has two types of fields in it.
Assume  $C$ and $D$ are scalar fields and $\beta$ and
$\bar{\beta}$ and fermionic fields in the superfield structure.
We have called the fermionic fields "ghosts" because they are
scalar fermions and also because they participate in the BRST
symmetry as we shall see below.

As the action is quadratic, the partition function can be
calculated explicitly

\begin{equation}\label{eq:a140}
Z=\int{\rm\bf D}C(x){\rm\bf D}D(x){\bf D}\bar{\beta}(x){\bf
D}\beta(x)e^{-S[C,D,\bar{\beta},\beta]}.
\end{equation}
The bosonic and fermionic parts of the action are decoupled and
integration over each of them can be performed independently. For
bosonic part one has
\begin{equation}\label{eq:a141}
Z_{b} = \int\prod_pdC_{p}\int\prod_pdD_p\exp \left\{-\frac{1}{2}
(\begin{array}{ll} C_p & D_p
\end{array})
G(p) \left(\begin{array}{l}
C_p \\
D_p
\end{array}\right)\right\},
\end{equation}
where a Fourier transform has been done and
\begin{equation}\label{eq:a142}
G(p)=\left(\begin{array}{cc}
  p^2+{m^2}_1 & 0 \\
  {m^2}_2 & p^2+{m^2}_1
\end{array}\right).
\end{equation}

These integrals are simple Gaussian ones. So, apart from some
unimportant numbers, this partition function becomes
\begin{equation}\label{eq:a143}
Z_b=\prod_p\left[\det\left(\begin{array}{cc}
  p^2+{m^2}_1 & 0 \\
  {m^2}_2 & p^2+{m^2}_1
\end{array}\right)\right]^{-1/2}=\prod_p(p^2+{m^2}_1)^{-1}.
\end{equation}
For the fermionic part the the same steps can be done and the
result is
\begin{equation}\label{eq:a144}
Z_f=\prod_p(p^2+{m^2}_1).
\end{equation}
Now it is easily seen that the total partition function is merely
a number, independent of the parameters of the theory and this is
the signature of BRST symmetry. Before proceeding, it is worth
mentioning that this symmetry will be induced onto LCFT part of
the correspondence and the correlation functions in that space
will also be invariant under proper transformations.

In BRST transformation, the fermionic and bosonic fields should
transform into each other. In our case this can be done using
$\eta$ and $\bar{\eta}$. Also one needs an infinitesimal
anticommuting parameters. Now let $\epsilon_1$ and $\epsilon_2$
be infinitesimal anticommuting parameters and consider the
following infinitesimal transformation of the the field $\Phi$
\begin{equation}\label{eq:a145}
\delta\Phi(\mathbf{x},\eta)=(\bar{\epsilon}\eta +
\bar{\eta}\epsilon) \Phi(\mathbf{x},\eta).
\end{equation}
It can be easily seen that this transformation leaves the action
invariant, because in the action the only terms which exist are
in the form of $\Phi(\eta)\Phi(-\eta)$ and under such a
transformation this term will become
\begin{equation}\label{eq:a146}
\delta(\Phi(\eta)\Phi(-\eta))=\Phi(\eta)(-\bar{\epsilon}\eta -
\bar{\eta}\epsilon) \Phi(-\eta)+(\bar{\epsilon}\eta +
\bar{\eta}\epsilon) \Phi(\eta)\Phi(-\eta),
\end{equation}
which is identically zero. We can therefore interpret this
transformation as the action of two charges $Q$ and $\bar{Q}$. The
explicit action of $Q$ for each component of $\Phi$ is
\begin{eqnarray}\label{eq:a147}
Q C&=&0,\nonumber\\
Q \beta&=&0,\nonumber\\
Q \bar{\beta}&=&C\nonumber\\
Q D&=&-\beta.
\end{eqnarray}
As expected the bosonic and fermionic fields are transformed into
each other, and the square of $Q$ vanishes. The action of
$\bar{Q}$ is similar except that it vanishes on $\bar{\beta}$ and
not on $\beta$.

To see how this symmetry is induced onto the LCFT part of the
theory one should first find the proper transformation. Going back
to equation (\ref{eq:a130}) and using the symmetry obtained for
the classical action one finds
\begin{equation}\label{eq:a151}
\exp(S_{Cl}[\Phi])= \left\langle\exp \left(\int O(\Phi+
\delta\Phi)\right) \right\rangle.
\end{equation}
As the transformation of $\Phi$ is $(\bar{\epsilon}\eta +
\bar{\eta}\epsilon)\Phi$ the integrand on the right hand side of
equation (\ref{eq:a151}) is just $O\Phi+(\bar{\epsilon}\eta +
\bar{\eta}\epsilon)O\Phi$ which can be regarded as $(O+\delta
O)\Phi$ with $\delta O=(\bar{\epsilon}\eta +
\bar{\eta}\epsilon)O$. So equation (\ref{eq:a151}) can be
rewritten as
\begin{equation}\label{eq:a152}
\left\langle\exp\left(\int O\Phi\right)\right\rangle=
\left\langle\exp\left(\int(O+\delta O) \Phi\right)\right\rangle.
\end{equation}
This shows that the correlation functions of the $O$ field are
invariant under the BRST transformation, that is $\delta\langle
O_1O_2\cdots O_n\rangle=0$ if the BRST transformation is taken to
be
\begin{equation}\label{eq:a153}
\delta O=(\bar{\epsilon}\eta + \bar{\eta}\epsilon)O.
\end{equation}
Again one can rewrite this transformation in terms of the
components of $O$ and the result is just the same as equation
(\ref{eq:a147}). This invariance can be tested using two point
correlation functions derived in previous section. These
correlation functions are easily found to be invariant under the
transformation law (\ref{eq:a153}), as an example
\begin{equation}\label{eq:a154}
Q \langle B \bar{\zeta}\rangle= \langle (Q B) \bar{\zeta}\rangle+
\langle B (Q \bar{\zeta}) \rangle= \langle\bar{\zeta}\zeta\rangle
- \langle BA\rangle=0.
\end{equation}
\section{Appendix: Hypergeometric functions}
In this appendix we show that equation (\ref{eq:a48}) can be
considered as 16 differential equations. However one of them is
trivial because it vanishes due to OPE constraints
\cite{mog,floh2}. According to equation (\ref{eq:a49}) $a$, $b$,
$c$ and so $H$ are functions of $\theta_{i}$'s. We write them in
a general form
\begin{eqnarray}\label{eq:a78}
H&=&\sum_{i=1}^{4}H_{i}\theta_{i}+\sum_{1\leq
i<j\leq4}H_{ij}\theta_{i}\theta_{j}+ \sum_{1\leq
i<j<k\leq4}H_{ijk}\theta_{i}\theta_{j}\theta_{k}+H_{1234}
\theta_{1}\theta_{2}\theta_{3}\theta_{4},\nonumber\\
a&=&a_{0}+\sum_{i=1}^{4}a_{i}\theta_{i}+\sum_{1\leq
i<j\leq4}a_{ij}\theta_{i}\theta_{j}+ \sum_{1\leq
i<j<k\leq4}a_{ijk}\theta_{i}\theta_{j}\theta_{k}+a_{1234}
\theta_{1}\theta_{2}\theta_{3}\theta_{4},
\end{eqnarray}
and in a similar way for $b$ and $c$. Now by substitution of them
in equation (\ref{eq:a48}) we obtain 15 differential equations
\begin{eqnarray}\label{eq:a79}
DH_{i}&=&0,\nonumber\\
DH_{ij}&=&\{-[c_{i}-(a_{i}+b_{i})\eta]
\frac{dH_{j}}{d\eta}+(a_{0}b_{i}+a_{i}b_{0})H_{j}+
i\longleftrightarrow j\},\nonumber\\
DH_{ijk}&=&[-[c_{k}-(a_{k}+b_{k})\eta]\frac{dH_{ij}}{d\eta}+
(a_{0}b_{k}+a_{k}b_{0})H_{ij}\nonumber\\
&-&[c_{ij}-(a_{ij}+b_{ij})\eta]\frac{dH_{k}}{d\eta}
+(a_{0}b_{ij}+a_{i}b_{j}+a_{ij}b_{0})H_{k}+cyclic\:\:\: terms
],\nonumber\\
DH_{1234}&=&[-[c_{l}-(a_{l}+b_{l})\eta]\frac{dH_{ijk}}{d\eta}+(a_{0}b_{l}+a_{l}b_{0})H_{ijk}\nonumber\\
&-&[c_{ij}-(a_{ij}+b_{ij})\eta]\frac{dH_{kl}}{d\eta}+(a_{0}b_{ij}+a_{i}b_{j}+a_{ij}b_{0})H_{kl}\nonumber\\
&-&[c_{ijk}-(a_{ijk}+b_{ijk})\eta]\frac{dH_{l}}{d\eta}+(a_{0}b_{ijk}+a_{ijk}b_{0}+a_{k}b_{ij}+a_{ij}b_{k})H_{l}\nonumber\\
&+&cyclic\:\:\:terms],
\end{eqnarray}
where
\begin{equation}\label{eq:a83}
D:=\eta(1-\eta)\frac{d^{2}}{d\eta^{2}}+[c_{0}-(a_{0}+b_{0}+1)\eta]
\frac{d}{d\eta}-a_{0}b_{0}.
\end{equation}

Let us now obtain from equation (\ref{eq:a51}), first few terms of
16 functions that appear in solutions of differential equations,
given above for the special case of $\Delta_{i}=\frac{1}{4}$
\begin{eqnarray}\label{eq:a84}
h_{0}&=& F(2,1,2,\eta)=1+\eta+\eta^{2}+\eta^{3}+\cdots\nonumber\\
h_{1}&=&\:\:\:\frac{1}{2}\eta+\frac{2}{3}\eta^{2}+\frac{3}{4}\eta^{3}+\cdots\:\:,\:\:\:\:\:\:\hspace{4mm}h_{2}=\frac{3}{2}\eta+\frac{7}{3}\eta^{2}+\frac{35}{12}\eta^{3}+\cdots\nonumber\\
h_{3}&=&-\frac{1}{2}\eta-\frac{2}{3}\eta^{2}-\frac{3}{4}\eta^{3}+\cdots\:\:,\:\:\:\:\:\:\hspace{4mm}h_{4}=\frac{1}{2}\eta+\frac{7}{9}\eta^{2}+\frac{35}{36}\eta^{3}+\cdots\nonumber\\
h_{12}&=&-\frac{1}{2}\eta-\frac{1}{18}\eta^{2}+\frac{29}{72}\eta^{3}+\cdots\:\:,\:\:\:\:\:\:h_{13}=\frac{1}{2}\eta+\frac{4}{9}\eta^{2}+\frac{3}{8}\eta^{3}+\cdots\nonumber\\
h_{23}&=&-\frac{2}{3}\eta^{2}-\frac{5}{4}\eta^{3}+\cdots\:\:,\:\:\:\:\:\:\hspace{13mm}
h_{14}=\frac{1}{3}\eta+\frac{2}{3}\eta^{2}+\frac{11}{12}\eta^{3}\nonumber\\
h_{24}&=&\frac{7}{6}\eta+\frac{70}{27}\eta^{2}+\frac{281}{72}\eta^{3}+\cdots\:\:,\:\:\hspace{1mm}h_{34}=-\frac{1}{2}\eta-\frac{49}{54}\eta^{2}-\frac{259}{216}\eta^{3}+\cdots\nonumber\\
h_{123}&=&\frac{7}{9}\eta^{2}+\frac{7}{6}\eta^{3}+\cdots\:\:,\:\:\:\:\:\:\hspace{15mm}h_{124}=-\frac{2}{3}\eta-\frac{41}{162}\eta^{2}+\frac{317}{648}\eta^{3}+\cdots\nonumber\\
h_{134}&=&\frac{2}{3}\eta+\frac{121}{162}\eta^{2}+\frac{455}{648}\eta^{3}+\cdots\:\:,\:\:\:\:h_{234}=-\frac{32}{27}\eta^{2}-\frac{46}{18}\eta^{3}+\cdots\nonumber\\
h_{1234}&=&\frac{137}{81}\eta^{2}+\frac{17}{6}\eta^{3}+\cdots.
\end{eqnarray}
where $Dh_{0}=0$.

\end{document}